\documentclass[aps,pra,reprint,showpacs,longbibliography,superscriptaddress]{revtex4-2}

\usepackage{xcolor}
\usepackage{amsmath}
\usepackage{transparent}
\usepackage{hyperref}
\usepackage{graphicx}
\usepackage{braket}
\usepackage{textcomp}
\usepackage{float}
\usepackage{soul} 
\usepackage{verbatim}
\usepackage{ulem}
\usepackage[per-mode=symbol, product-units=single, separate-uncertainty=true]{siunitx}
\usepackage{xurl}

\begin{document}

\title{Suppressing the Motion of Rydberg Atoms in Inhomogeneous Electric Fields via Stark Echo}
\author{Dominik Jakab}\email{dominik.jakab@uni-tuebingen.de}
\affiliation{Center for Quantum Science, Physikalisches Institut, Eberhard Karls Universit\"at
	T\"ubingen, Auf der Morgenstelle 14, D-72076 T\"ubingen, Germany}
\author{Manuel Kaiser}
\affiliation{Center for Quantum Science, Physikalisches Institut, Eberhard Karls Universit\"at
	T\"ubingen, Auf der Morgenstelle 14, D-72076 T\"ubingen, Germany}
\author{Conny Glaser}
\affiliation{Center for Quantum Science, Physikalisches Institut, Eberhard Karls Universit\"at
	T\"ubingen, Auf der Morgenstelle 14, D-72076 T\"ubingen, Germany}
\author{David Petrosyan}
\affiliation{Center for Quantum Science, Physikalisches Institut, Eberhard Karls Universit\"at
	T\"ubingen, Auf der Morgenstelle 14, D-72076 T\"ubingen, Germany}
\affiliation{Institute of Electronic Structure and Laser and Center for Quantum Science and Technologies, FORTH, 70013 Heraklion, Crete, Greece}
\author{József Fortágh}
\affiliation{Center for Quantum Science, Physikalisches Institut, Eberhard Karls Universit\"at
	T\"ubingen, Auf der Morgenstelle 14, D-72076 T\"ubingen, Germany}
\author{Andreas Günther}
\affiliation{Center for Quantum Science, Physikalisches Institut, Eberhard Karls Universit\"at
	T\"ubingen, Auf der Morgenstelle 14, D-72076 T\"ubingen, Germany}
	
\date{\today}

\begin{abstract}
    Rydberg atoms possess strong electric dipole transitions and tunable energy levels, making them promising candidates for microwave to optical conversion on integrated superconducting atom chips. Achieving strong coupling of the atoms to e.g. the microwave field of an on-chip resonator requires placing the atoms within tens of micrometers from the chip surface. However, inhomogeneous stray electric fields originating from the surface can induce position-dependent Stark forces, resulting in atomic motion and leading to time-dependent shifts of the Rydberg energy levels. We experimentally investigate these effects using time-of-flight and spectroscopic techniques, observing substantial level shifts and signal loss attributable to field-induced atomic motion. A theoretical model incorporating an exponentially decaying surface field with a superimposed bias accurately reproduces the observed dynamics. To mitigate the level shift, we introduce a Stark echo sequence that dynamically reverses the force. This approach suppresses the atomic motion and maintains the atomic resonance. The method relies solely on global field control and is compatible with atom–resonator coupling architectures, providing a robust strategy for preserving coherence of Rydberg atoms in inhomogeneous electric fields near surfaces.

\end{abstract}

\maketitle

\section{Introduction}\label{sec:introduction}
    A promising route towards microwave-to-optical quantum transduction leverages the strong electric-dipole transitions between Rydberg states of atoms \cite{han2018coherent,kumar2023quantum,Borówka_Pylypenko_Mazelanik_Parniak_2024}. The atoms can be coupled to high-Q superconducting resonators \cite{Morgan_Hogan_2020}, which can also host superconducting qubits \cite{blais2021circuit} and can be conveniently integrated on superconducting atom chips \cite{Kaiser} that allow trapping ultracold ensembles nearby. Achieving strong atom–resonator coupling requires placing the atoms within tens of micrometers of the chip surface, where the microwave field is sufficiently strong \cite{Petrosyan_Mølmer_Fortágh_Saffman_2019}.
    However, at these distances, stray electric fields originating from surface adsorbates, patch potentials, trapped charges, or material inhomogeneities are a significant source of dephasing \cite{Hattermann_Mack_Karlewski_Jessen_Cano_Fortágh_2012, Carter_Martin_2011, Abel_Carr_Krohn_Adams_2011}. These fields are typically spatially inhomogeneous, leading to position-dependent Stark shifts and forces acting on the Rydberg atoms.

    Several strategies have been explored to suppress stray electric fields near surfaces, including active compensation using electrodes \cite{Panja_Wang_Wang_Wang_Subhankar_Liang_2024, Glaser_IonOptics}, ultraviolet (UV)-induced desorption of surface contaminants \cite{Davtyan_2018, Ocola_2024}, and the application of specialized surface coatings \cite{Hermann_2014}. While these methods can significantly reduce field amplitudes, complete cancellation of the field remains challenging. This is due to the presence of multiple, only partially characterized, and often time-dependent sources of electric fields. Consequently, residual field gradients frequently persist even in carefully compensated systems. These gradients produce spatially varying Stark shifts and can exert substantial forces on the atoms \cite{Hogan_2016}, driving motional dynamics and inducing spatially- and temporally varying Rydberg level shifts. Such effects can be the major source of dephasing and coherence loss in atom–resonator coupling schemes and consequently limit the timescale for cavity-based microwave to optical conversion \cite{Petrosyan_Mølmer_Fortágh_Saffman_2019} and atom-atom quantum gates \cite{Sárkány_Fortágh_Petrosyan_2018} to a few microseconds or less.

    In this work, we report on time-dependent resonance shifts and reduced lifetimes of Rydberg atoms in an ultracold atomic cloud positioned near a superconducting chip surface, consistent with motional dynamics induced by spatially inhomogeneous electric fields. We develop a model in which atoms experience Stark-induced forces arising from the gradient of an exponentially decaying surface field superimposed with a controllable bias. Numerical simulations based on this model accurately reproduce the observed energy shifts and atomic displacement over time. To mitigate the resulting level shifts, we implement a Stark echo sequence that dynamically reverses the force by switching the bias field amplitude. This technique effectively suppresses the motion, maintaining the atomic resonance condition for durations exceeding \qty{10}{\us}. The method relies solely on global field control and is experimentally validated in our chip-based Rydberg atom platform.
    Our approach offers a robust and scalable solution for maintaining the atomic resonance conditions in hybrid atom–chip architectures, with direct relevance to quantum transduction and other applications in quantum computation with Rydberg atoms.

\section{Theoretical model}\label{sec:ResShift}
 \subsection{Electric field–induced atomic motion and level shift}
   Atoms in Rydberg states exhibit exceptionally large electric polarizabilities, scaling approximately as $\alpha \propto n^7$ with principal quantum number $n$ \cite{gallagher2006rydberg}. This makes them highly sensitive to external electric fields. In inhomogeneous fields, this sensitivity leads not only to static Stark shifts but also to spatially dependent forces that accelerate the atoms. The resulting spatially and temporally varying energy shifts can, for example, detune the atoms with respect to a fixed cavity resonance frequency and lead to dephasing of the atom-cavity interaction. 
   
   For low-angular-momentum states in small electric fields, the Stark shift $\mathcal{E}(\vec{E})$ is well described by \cite{PhysRevA.3.1209}
    \begin{equation}\label{eq:potential}
        \mathcal{E}(\vec{E}) = -\frac{1}{2} \alpha_{\mathrm{0}} E^2,
    \end{equation}
    where $\alpha_{\mathrm{0}}$ is the scalar polarizability and we set the quantization axis $z$ along the electric field direction $\vec{E} = E \vec{e}_z$. For simplicity, we neglected the tensor polarizability, as it is vanishing for $S$ states in alkali atoms and generally small for low-angular-momentum states \cite{Yerokhin_Buhmann_Fritzsche_Surzhykov_2016}. In a static electric field, the Rydberg atom acquires an induced dipole moment $\mu(E) = - \partial \mathcal{E}(E) / \partial E = \alpha_0 E$.
    
    To compute the polarizability $\alpha_{\mathrm{0}}$ and induced dipole moment $\mu(E)$, we take the Stark Hamiltonian $\hat{H} = \hat{H}_0 - e E\hat{z}$ of a single atom in the field-free eigenbasis of $\hat{H}_0$, then numerically diagonalize it to calculate the eigenvalues of the perturbed Hamiltonian $\hat{H}$ for different electric field strengths \cite{zimmerman1979stark, grimmel2015measurement}. For the $42S_{1/2}$ state of Rubidium 87 used in this study, the Stark shift remains approximately quadratic over the relevant field range of a few \unit{\V \per \cm}, such that Eq.~(\ref{eq:potential}) yields a good approximation with $\alpha_{\mathrm{0}}/(2\pi\hbar) \approx \qty{15}{\MHz \per (\V/\cm)^2}$ and $\hbar$ being the reduced Planck constant.
    
    An atom at position $z$ experiences a force along the $z$-axis in a spatially varying field $E(z)$:
    \begin{equation}\label{eq:starkforce}
        \vec{F}(z) = F(z)\vec{e}_z = - \vec{\nabla} \mathcal{E} = \alpha_{\mathrm{0}} E(z) \frac{dE(z)}{dz}\vec{e}_z.
    \end{equation}
    In experiments with atoms trapped close to (chip) surfaces, the electric field is strongly inhomogeneous perpendicular to the surface. We model this field as an exponentially decaying surface field $E_{\mathrm{sur}}$  superimposed with a tunable homogeneous bias field $E_{\mathrm{bias}}$  \cite{Hattermann_Mack_Karlewski_Jessen_Cano_Fortágh_2012}
    \begin{equation}
        E(z) = E_\text{sur}(z) + E_{\mathrm{bias}} = E_0 e^{-z/\zeta} + E_{\mathrm{bias}},
        \label{eq:totalField}
    \end{equation}
    where $E_0$ is the field amplitude at the surface and $\zeta$ the characteristic decay length. This yields a position-dependent force
    \begin{equation}
        F(z) = - \frac{\alpha_{\mathrm{0}} E_0}{\zeta} e^{-z/\zeta} \left(E_0 e^{-z/\zeta} + E_{\mathrm{bias}} \right).
        \label{eq:force}
    \end{equation}
    
    \begin{figure}
        \centering
        \includegraphics[]{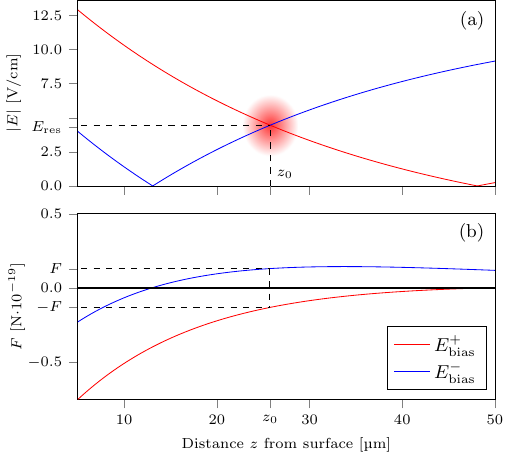}
        \caption{Electric field and resulting Stark force near the chip surface.
            (a) Magnitude of the total electric field $|E(z)|$ as a function of distance $z$ from the chip surface, including the exponentially decaying surface field and the tunable bias. Red and blue lines correspond to $E_{\mathrm{bias}} = \qty{-4.02}{\V \per \cm}$ and $\qty{-12.93}{\V \per \cm}$, respectively. At the position of the atomic cloud ($z_0 = \qty{25.75}{\um}$), both values of the bias field produce the same total field magnitude ($|E_\text{res}| = \qty{4.45}{\V \per \cm}$), but with field gradients of opposite sign. The red shaded area illustrates the spatial extent of the cloud.
            (b) Stark force $F(z)$ experienced by ${}^{87}$Rb atoms in the $42S_{1/2}$ Rydberg state, calculated for the above field with $\alpha_0/(2\pi\hbar) \approx  \qty{15}{\MHz \per (\V/\cm)^2}$, given by Eq.~(\ref{eq:starkforce}). At $z_0 = \qty{25.75}{\um}$, the force reverses sign between the two configurations, accelerating atoms either toward (red) or away from (blue) the chip surface. Electric field parameters: $E_0 = \qty{20}{\V \per \cm}$, $\zeta = \qty{30}{\um}$.}
        \label{fig:TotalFieldBias}
    \end{figure}
    
    We denote the unshifted (zero-field) transition frequency from the ground state to Rydberg state $\ket{r}$ as $\omega_r(0)$. In a static electric field the transition frequency is shifted according to Eq.~(\ref{eq:potential}) to $\omega_r(E) = \omega_r(0) - \alpha_0 E^2/(2\hbar)$, where the ground state Stark shift is neglected. The Rydberg atom is excited via a laser with a fixed frequency $\omega_\text{exc} = \omega_r(0) - \Delta \omega_\text{exc}$, where $\Delta \omega_\text{exc}$ is the laser detuning with respect to the zero-field resonance. The electric field $E_\text{res}$ that tunes the atom to resonance with the laser must satisfy $\omega_\text{exc} = \omega_r(E_\text{res}) = \omega_r(0) - \alpha_0 E_\text{res}^2/(2\hbar)$, so that
    \begin{equation}\label{eq:res_field}
        E_\text{res} = \sqrt{2 \Delta \omega_\text{exc} \hbar / \alpha_0}.
    \end{equation}
     A single atom in a thermal cloud at position $z_0$ can then be resonantly excited only if $|E(z_0)| = E_\text{res}$. Combining with Eq.~(\ref{eq:totalField}), this leads to the condition for the bias field:
    \begin{equation}\label{eq:bias_fields}
        E_{\text{bias}}^\pm = \pm E_\mathrm{res} - E_\mathrm{sur}(z_0) = \pm \sqrt{\frac{2 \Delta \omega_\text{exc}\hbar}{\alpha_0}} - E_0 e^{-z_0/\zeta}.
    \end{equation}
    According to Eq.~(\ref{eq:force}), these bias fields $E_{\text{bias}}^\pm$ lead to two forces of equal magnitude but opposite directions at position $z_0$:
    \begin{equation}
        F_{\mp} = \mp \frac{\alpha_{\mathrm{0}} E_0 E_\mathrm{res}}{\zeta} e^{-z_0/\zeta} = \mp \frac{E_\text{sur}(z_0) \mu(E_\text{res})}{\zeta}.
    \end{equation}
    Figure~\ref{fig:TotalFieldBias} illustrates the total field magnitude and forces for the two bias field values.
    
    In summary, the Rydberg atom experiences either an attractive or repulsive force from the surface and will move towards or away from it. Due to this motion in the inhomogeneous field, the atom at position $z(t)$ will experience a time-dependent detuning $\Delta \omega_r (t)$ from the excitation laser frequency, given by
    \begin{eqnarray}\label{eq:energy_shift}
        \Delta \omega_r (t) &=& \omega_r(E(z(t))) - \omega_\text{exc} \\
                            &=& -\frac{1}{\hbar}\left(\mu(E_\text{res})+\frac{1}{2}\mu(\Delta E)\right)\Delta E,\nonumber
    \end{eqnarray}
    where $\Delta E = E(z(t))-E(z_0)$ and $E_\text{res} = E(z_0)$. The magnitude of this shift is determined by the (surface) electric field and the induced dipole moment $\mu$. While it would be possible to set the total electric field to zero at the cloud center to minimize the force, the resulting equilibrium at $z_0$ would be unstable for Rydberg states with $\mu > 0$. Moreover, a finite electric field provides an easy and precise method for tuning of Rydberg transitions. When an external cavity resonance cannot be tuned over a large frequency range (e.g. in a coplanar waveguide (CPW) resonator structure, as introduced in the next section), the atomic resonance must be tuned to the cavity mode. This necessitates the use of non-zero electric fields at the cloud position and motivates the use of echo techniques discussed below. 
        
\subsection{Stark echo for motion reversal}\label{sec:EchoSim}

    The atomic motion in the inhomogeneous electric field and therefore the time-dependent Rydberg level shift can be suppressed by a pulse sequence resembling the well-known spin echo sequence. The idea is as follows: the sign of the total electric field at atomic position $z_0$ — and thus the direction of the Stark force — is flipped after a given time-of-flight (TOF) $t_\text{flip}$. This effectively reverses acceleration and returns the atoms to their initial position at the echo time $t_\text{echo}$. In principle, the reversal can be done repetitively, such that the sign of the force is periodically flipped during the total time evolution of the system. The atoms then oscillate around their initial position and the average level shift remains approximately constant.
    
    First, we consider a single echo sequence of total length $t_\text{echo}$ with a field flip at $t_\text{flip}<t_\text{echo}$. Let us denote the displacement of the atom with mass $m$ after time $t$ by $s(t)$. Then, as the atom shall be back at $z_0$ after $t_\text{echo}$, we have $s(t_\text{echo}) = 0$. From the kinematic equations we thus have
    \begin{equation}\label{eq:kinematic}
        s(t_\text{echo}) = \frac{\left|F_{\mp}\right|}{2 m} t_\text{flip}^2 + \left( v_\text{flip} \tau - \frac{\left|F_{\mp}\right|}{2 m} \tau^2 \right) = 0,
    \end{equation}
    where the first and second terms are the displacement before and after the first flip, respectively, $\tau = t_\text{echo} - t_\text{flip}$, and $v_\text{flip} = \left|F_{\mp}\right| t_\text{flip} / m$ is the velocity at $t_\text{flip}$. Here, we neglected the position dependence of $F$ as the atom can be considered localised in the vicinity of $z_0$ if $t_\text{echo}$ is sufficiently short. Solving Eq.~(\ref{eq:kinematic}) and disregarding the negative solution yields $t_\text{flip} = ((\sqrt{2}-1) / \sqrt{2}) t_\text{echo} \approx 0.3 t_\text{echo}$ and $\tau\approx 0.7 t_\text{echo}$. The sequence $t_\text{flip} - \tau$ with a field flip at $t_\text{flip}$ thus concludes the single echo sequence. 
    
    We can extend the scheme for a sequence of echo pulses. After the first pulse, the atoms will cross $z_0$ at $t=t_\text{echo}$ with some final velocity $v_\text{echo} = \left|F_{\mp}\right| \left(t_\text{flip} - \tau\right) / m$. From here on the oscillation dynamics can be periodically repeated by flipping the field each time the atoms cross $z_0$. This happens at constant time intervals $t_{\text{cyc}}$ with $s(t_\text{echo} + nt_\text{cyc}) = 0$, where $n \geq 0$ is an integer number. From the kinematic equation we find
    \begin{equation}
            t_\text{cyc} = \frac{2(2-\sqrt{2})}{\sqrt{2}}t_{\text{echo}} \approx 0.83t_{\text{echo}}.
    \label{eq:flip_timings}
    \end{equation} 
    
    We note, that both the single-pulse ($n=0$) and multi-pulse ($n \geq 1$) echo sequences allow to retrieve the original position at times $t_n = t_\text{echo} + n t_\text{cyc}$. However, the velocity is $v_n = (-1)^{n+1} v_\text{echo}$. If required an additional flip may be added at the beginning or end of the pulse sequence to also retrieve the original field conditions. Other more complex schemes such as a $t_\text{flip} - \tau - \tau - t_\text{flip}$ scheme or a $T - 2T - T$ scheme with arbitrary $T$ may be used to retrieve position and velocity at the same time. However, for sake of simplicity we limit ourselves to the aforementioned $t_\text{flip} - \tau - t_\text{cyc}$ scheme.

\begin{figure}
        \centering
        \includegraphics[]{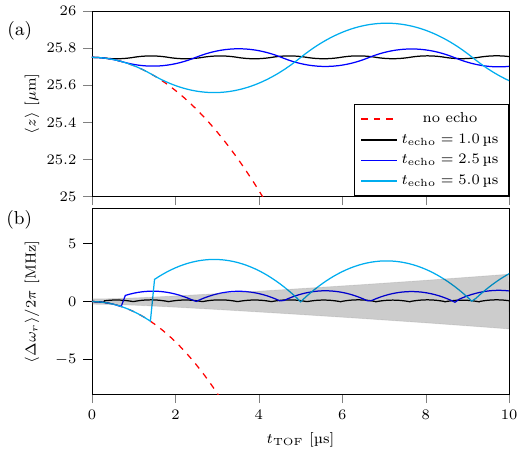}
        \caption{Numerical simulation of Stark echo suppression of motional level shift.
            (a) Simulated center-of-mass position $\langle z \rangle$ as a function of $t_{\text{TOF}}$ for atoms excited from an atomic cloud centered at $z_0 = \qty{25.75}{\um}$ with temperature $T = \qty{1}{\micro \K}$. The solid black, blue and cyan lines show the results using Stark echo with $t_\mathrm{echo} = \qty{1}{\us}$, \qty{2.5}{\us} and \qty{5}{\us}, respectively. Due to the echo sequence, the atoms oscillate around their initial position, with the oscillation amplitude depending on $t_\mathrm{echo}$. The dashed red line shows the trajectory without the echo sequence, with the atoms rapidly accelerating away from their initial position.
            (b) Corresponding mean level shift $\langle \Delta \omega_r \rangle$ of $42S_{1/2}$ Rydberg atoms. The solid black, blue and cyan lines show the simulated mean shift for the Stark echo durations mentioned above. As $t_\mathrm{echo} \longrightarrow 0$, the residual oscillations of the level shift are suppressed due to the stronger spatial confinement. The gray shaded area indicates the $1\sigma$ standard deviation of the individual atomic shifts $\{ \Delta\omega_r^i\}$ for $t_\mathrm{echo} = \qty{1}{\us}$ and the red dashed line shows the simulated shift without the echo sequence. }
        \label{fig:EchoSim}
    \end{figure}

    We simulate the dynamics for individual ${}^{87}$Rb atoms in the $42S_{1/2}$ state sampled from a thermal cloud at $T=\qty{1}{\micro K}$, with electric field parameters $E_0 = \qty{20}{\V \per \cm}$ and $\zeta = \qty{30}{\um}$. Each Rydberg atom at position $z$ is taken from a region of a Gaussian cloud centered around $z_0=\qty{25.75}{\um}$ where the field is approximately resonant $|E(z) - E_\text{res}| < \qty{5}{\mV \per \cm}$, with cloud parameters given in Sec.~\ref{subsec:setup}. 
    After numerically solving the equations of motion for the atoms, the resulting level shift is extracted for each atom $i$ as $\Delta \omega_r^i$, and averaged over $N_\text{sim} = 1000$ atoms $\langle \Delta \omega_r \rangle = \sum_{i=1}^{N_\text{sim}} \Delta \omega_r^i / N_\text{sim}$. The result is shown in Fig.~\ref{fig:EchoSim}(a,b). For an echo time of $t_\text{echo}=\qty{1}{\micro\second}$ (solid black line), the atoms are spatially confined to $z_0 \pm \qty{10}{\nm}$ (a) and the ensemble-averaged shift $\langle \Delta \omega_r \rangle / 2 \pi$ remains below $\qty{0.15}{\mega\hertz}$ (b) for more than \qty{10}{\micro\second}, demonstrating the effectiveness of the method for maintaining atomic resonance. Without the echo sequence (red dashed line), the atoms are quickly accelerated away and experience a large level shift. We also simulated the motional dynamics for different echo pulse lengths of $t_\text{echo}=\qty{2.5}{\micro\second}$ and \qty{5}{\micro\second}, shown with the blue and cyan lines, respectively. These results indicate that as $t_\text{echo}\longrightarrow 0$, the spatial confinement gets stronger, which leads to smaller oscillations in the mean level shift. 
    
    If $t_{\text{TOF}} \gg \qty{10}{\us}$ (or the electric field is strongly inhomogeneous), the spatial dependence of $F(z)$ (see Eq~(\ref{eq:starkforce})) cannot be neglected anymore. This would introduce a small inbalance of the force around $z_0$, which leads to a slow drift of the atomic position over time. As $t_\text{echo}\longrightarrow 0$, this effect is also strongly suppressed, as due to the spatial confinement the atoms explore less of the inhomogeneous electric field landscape.
    
    We analyzed the influence of the thermal motion of the atoms and found that for temperatures in the range of $T=\qtyrange{1}{100}{\mu K}$ the mean level shift is not significantly affected, but with increasing temperature the atoms experience a strong broadening of the transition resonance $\sigma (\Delta \omega_r)$, defined as the standard deviation of the individual atomic shifts $\{ \Delta\omega_r^i\}$. This is due to the thermal expansion also exposing the atoms to different fields in the inhomogeneous electric field landscape. Surprisingly, the echo sequence reduces the broadening by a factor of $\sim 1.6$. The broadening due to thermal motion over \qty{10}{\us} is $\sigma(\Delta \omega_r) / 2 \pi \approx \qty{4.93}{\MHz}$, $\qty{15.28}{\MHz}$ and $\qty{50.89}{\MHz}$ for \qty{1}{\micro \K}, \qty{10}{\micro \K} and \qty{100}{\micro \K}, respectively, and for \qty{1}{\micro \K} we show it as shaded gray region in Fig.~\ref{fig:EchoSim}(a).  Without the echo sequence, the broadenings would be $\qty{8.34}{\MHz}$, $\qty{24.96}{\MHz}$ and $\qty{79.19}{\MHz}$, respectively.
    
    In realistic implementations, imperfections such as finite rise times of voltage ramps limiting $t_\text{echo}$, and electrical noise reduce the cancellation efficiency and increase the residual oscillations. Although we focus on field inhomogeneities in the z-direction, the concept of motional echo can, in principle, be extended to suppress motion in any directions. Lateral ($x/y$) electric field components are typically also present, for example due to surface contaminants or patch potentials \cite{obrecht2007measuring}. With proper control over the homogeneous bias field in all three directions, it would be possible to locally reverse the gradient and thereby reverse the motion in any direction.
    
\section{Experimental Implementation and Results}\label{sec:EchoExp}

    \subsection{Experimental setup}\label{subsec:setup}
    
        \begin{figure}
            \centering
            \includegraphics[width=0.5\textwidth]{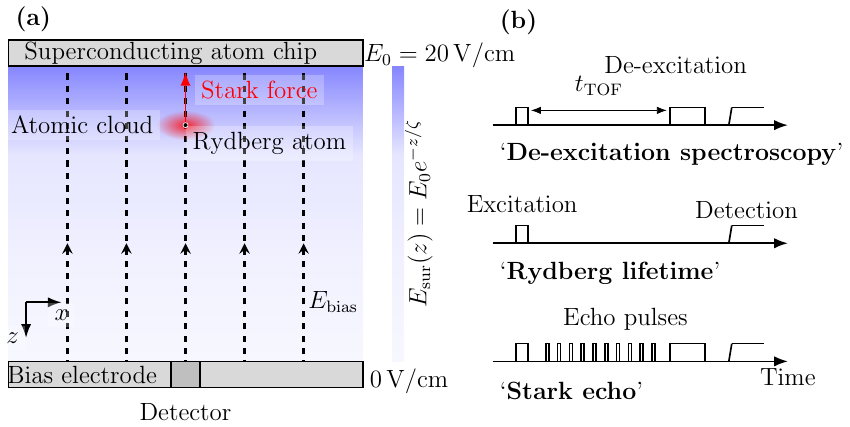}
            \caption{Schematics of the experiment. (a) A cloud of atoms (red) is magnetically trapped at a distance of $z_0\approx$ \qty{25}{\um} below a superconducting atom chip. The atoms are located in a stray electric field (blue shaded gradient) decaying exponentially with distance from the chip surface, $E_{\text{sur}}(z) = E_0 \exp(-z/\zeta)$, and a homogeneous bias field $E_\text{bias}$ (black dashed arrows) is generated by an electrode placed \qty{4.5}{mm} below the chip surface, forming a plate capacitor with the chip. The inhomogeneous total electric field produces a position-dependent Stark force (red arrow), which accelerates the atoms and shifts the Rydberg energy level.
            (b) Pulse sequences for the three measurement protocols. Top: de-excitation spectroscopy: atoms are excited to a Rydberg state at a given detuning $\Delta\omega_\text{exc}$, evolve for some $t_\text{TOF}$, then de-excited by another laser and finally detected via field ionization. Middle: Lifetime measurement: atoms are excited and detected after variable $t_{\text{TOF}}$ without de-excitation. Bottom: Stark echo sequence: electric field gradient is dynamically inverted (echo pulses) during TOF, reversing the Stark-induced acceleration and level shift.}
            \label{fig:Scheme}
        \end{figure}
        
        A cloud of approximately $\num{1.5e5}$ ${}^{87}$Rb atoms is prepared in the $5S_{1/2} \ket{F=2,m_F=2}$ ground state and magnetically trapped $\sim \qty{25}{\um}$ below a superconducting atom chip surface, in the vicinity of a CPW resonator as described elsewhere \cite{Hattermann}. The cloud is cooled to about \qty{1}{\micro\kelvin} temperature and has an elongated shape with full widths at half maximum (FWHM) of $\qty{100}{\um}$, $\qty{10}{\um}$, $\qty{10}{\um}$ along the $x$-, $y$-, $z$-direction, respectively.
    
        Rydberg excitation to the $42S_{1/2}$ state is performed in a two-photon scheme via the intermediate $6P_{3/2}$ state using counter-propagating beams of lasers with wavelengths \qty{420}{\nm} and \qty{1020}{\nm}. The blue beam has a waist of $\sim \qty{2}{\mm}$ and power of \qty{30}{\micro\watt}, while the infrared beam is focused to a waist of $\sim \qty{15}{\um}$ with the power of \qty{60}{\milli\watt}. Both lasers are frequency-stabilized to a ULE reference cavity with a linewidth $< \qty{10}{\kHz}$ and detuned with respect to the intermediate state by $\pm\qty{40}{\MHz}$. Laser pulses are generated using acousto-optic modulators, with excitation pulse durations of \qty{0.5}{\us}. The power is adjusted such that on average one Rydberg atom is detected per excitation pulse. The pulse repetition rate is \qty{2}{kHz} and a measurement cycle contains $N_\text{rep}=1000$ pulses for each cloud.
        
        Close to the chip surface ($\approx \qty{25}{\um}$), stray electric fields on the order of \qty{10}{V/cm} are observed. These fields remain stable over the course of the measurements (several hours) and are characterized using Rydberg atom spectroscopy, as described in \cite{Kaiser}. From these measurements, we infer an exponentially decaying field strength perpendicular to the chip surface, as per Eq.~(\ref{eq:totalField}), with a surface field amplitude $E_0\approx \qty{20}{\V \per \cm}$ and a characteristic decay length $\zeta \approx \qty{30}{\um}$. Tuning of the vertical electric field ($z$) component is achieved through a large bias electrode, which provides a homogeneous bias field $E_\text{bias}$ along the $z$-axis, with the chip held at ground potential. The resulting total field, including stray surface fields and applied compensation, is described by Eq.~(\ref{eq:totalField}). The experimental setup including the thermal cloud, electrodes and electric field structure is illustrated in Fig.~\ref{fig:Scheme}(a).
        
        After excitation, the atoms evolve in the combined electric field for a given free evolution time $t_\text{TOF}$ (see Fig.~\ref{fig:Scheme}(b), and are subsequently detected, either directly or following de-excitation induced by a \qty{1020}{\nm} laser pulse.
        Detection is performed via field ionization: the Rydberg atoms are ionized by a $\sim\qty{1}{\us}$ voltage ramp applied to the bias/extraction electrode located \qty{4.5}{\mm} below the atom chip surface (see Fig.~\ref{fig:Scheme}). The resulting electrons are extracted through a central bore in the electrode and detected by a channel electron multiplier. The complete charged particle optics system is described in detail elsewhere \cite{Glaser_IonOptics}. The pulse sequences containing the excitation, time evolution, de-excitation, and ionization are illustrated in Fig.~\ref{fig:Scheme}(b).
        
    \subsection{Rydberg level shift and lifetime} \label{subsec:detuning_lifetime}
        To characterize the effect of inhomogeneous surface fields on the atomic motion, we perform de-excitation spectroscopy at varying times after Rydberg excitation. Atoms are excited to the $42S_{1/2}$ state and, after a variable time-of-flight, de-excited using a \qty{1.0}{\us} laser pulse at frequency $\omega_\text{de-exc}$. As the detection scheme is limited to Rydberg atoms, resonant de-excitation is observed as a measurable reduction in the ionization signal. The de-excitation laser detuning from the excitation frequency $\Delta \omega_\text{de-exc} = \omega_\text{de-exc} - \omega_\text{exc}$ is scanned and the resulting spectra is fitted by an inverted Gaussian profile, from which we extract the resonance position $\langle \Delta \omega_r \rangle$ and the full width at half maximum (FWHM).

        \begin{figure}
            \centering
            \includegraphics[]{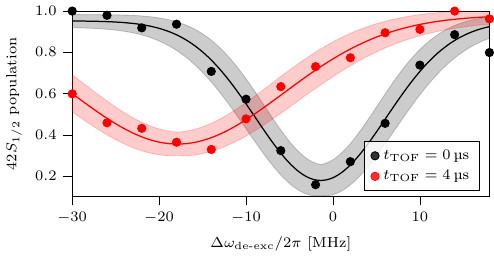}
            \caption{De-excitation spectra of Rydberg atoms measured for $\Delta\omega_\text{exc}/2\pi\approx\qty{150}{\MHz}$ and different $t_{\text{TOF}}$ after excitation. Black circles: Spectrum for $t_\text{TOF} = 0$, with an inverted Gaussian fit (solid line) yielding a detuning of $\Delta \omega_\text{de-exc} / 2\pi = \qty{-1.4}{\MHz}$ and a FWHM of \qty{17.7}{\MHz}. Red circle: Spectrum after $t_\text{TOF} = \qty{4}{\us}$, showing a pronounced red shift ($\Delta \omega_\text{de-exc} / 2\pi = \qty{-17.9}{\MHz}$) and spectral broadening ($\text{FWHM} = \qty{28.7}{\MHz}$). Each data point is averaged over $N_\text{rep} = 1000$ pulse sequences. The error band denotes the $1\sigma$ confidence interval of the fit.}
            \label{fig:de-excitationSpectrum}
        \end{figure}
        
        Figure~\ref{fig:de-excitationSpectrum} shows the de-excitation spectra for two representative $t_\text{TOF}$ values. For $t_\text{TOF} = 0$, the Rydberg level is mostly unshifted with $\langle \Delta \omega_r \rangle /2\pi \approx \qty{-1.4}{MHz}$. After \qty{4}{\us}, however, the spectrum is significantly red-shifted (by $\sim \qty{18}{\MHz}$) and broadened. This indicates both a net shift of the Rydberg energy level (the Stark shift of the 6P state is negligible) and an increased spread of the level shifts due to position-dependent forces and thermal motion in the inhomogeneous electric field. 
        
\begin{figure}
            \centering
            \includegraphics[]{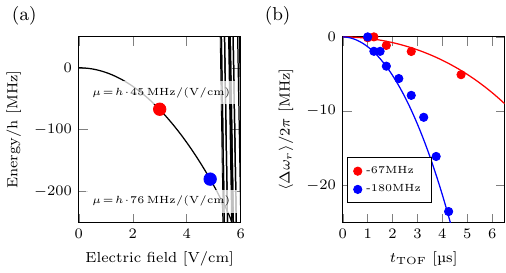}
            \caption{(a) Simulated Stark map for the Rydberg state $42S_{1/2}$, taking the field-free energy level $\omega_r(0)$ as zero, and illustrating the dependence of the induced dipole moment $\mu(E)$ on $E$ as the potential gradient.
            (b) Measured shift of de-excitation resonance as a function of time-of-flight for Rydberg atoms excited at two different detunings from the zero-field resonance ($\Delta\omega_\text{exc}/2\pi$=\qty{67}{\MHz} and \qty{180}{\MHz}). Atoms with larger initial detuning possess larger induced dipole moments and experience stronger acceleration in the inhomogeneous electric field, resulting in a larger level shift over time. The solid lines represent numerical simulations using measured electric field parameters stated in Sec~\ref{subsec:setup}.}
            \label{fig:StarkMap}
        \end{figure}

        To better understand the origin of the observed level shift, we repeat the measurement with a larger excitation laser detuning $\Delta \omega_\text{exc}$, effectively increasing $E_\text{res}$ (cf. Eq.~\ref{eq:res_field}). This, in turn, increases the dipole moment as shown in Fig~\ref{fig:StarkMap}(a), leading to a larger level shift according to Eq.~(\ref{eq:energy_shift}). For each detuning $\Delta \omega_\text{exc}$, we adjust the bias field $E_{\text{bias}}$ to ensure that the atoms located at the cloud center $z_0$ are resonant with the excitation.
        Figure~\ref{fig:StarkMap}(b) shows the measured de-excitation resonance positions as a function of $t_{\text{TOF}}$ for two excitation detunings. Each data point corresponds to a full de-excitation spectrum (as in Fig.~\ref{fig:de-excitationSpectrum}), with the resonance position extracted from the Gaussian fit. For an excitation laser detuning of $\Delta \omega_\text{exc}/2\pi=\qty{67}{\MHz}$ (red points), the atoms possess an induced dipole moment of approximately $h\cdot(45\,\mathrm{MHz}/(\mathrm{V}/\mathrm{cm}))$, where $h$ is the Planck constant, as seen from the gradient in the simulated Stark map in Fig.~\ref{fig:StarkMap}(a). For the larger detuning of \qty{180}{\MHz} (blue points), the corresponding dipole moment is larger, around $h\cdot(76\,\mathrm{MHz}/(\mathrm{V}/\mathrm{cm}))$.
        This difference in dipole moments results in significantly different motional behavior. At $t_{\text{TOF}} = \qty{4}{\us}$, the atoms with stronger dipole moment (blue) exhibit a level shift of more than $\qty{-21}{\MHz}$, while the atoms with weaker dipole moment (red) experience a 7-fold smaller shift of $\sim \qty{-3}{\MHz}$. 
        Our measurements show, that the Rydberg energy level shift depends on the excitation frequency and is quadratic in time, indicating the acceleration of the atoms under a position-dependent force scaling with the dipole moment.
        
        To confirm these observations, we compare the experimental data with 3D numerical calculations for two different excitation frequencies/dipole moments. The solid red and blue lines in Fig.~\ref{fig:StarkMap}(b) show the simulated time-dependent level shift caused by the atomic motion in the field gradient together with the measured level shift. The \textit{ab initio} calculations fit the experimental data well. Based on these simulations, even for the moderately polarizable $42S_{1/2}$ state, atoms acquire frequency shifts exceeding $\qty{100}{\MHz}$ within \qty{10}{\us} for $\Delta \omega_\text{exc}/2\pi=\qty{180}{\MHz}$.

        In a second measurement, we investigate how atomic motion in the electric field affects the Rydberg state population itself. Therefore we measure the remaining Rydberg population as a function of $t_{\text{TOF}}$ without applying a de-excitation pulse (cf. Fig.~\ref{fig:Scheme}(b), middle plot). Here, atoms are excited at a total electric field magnitude of $|E_\text{res}| = \qty{4.45}{\V \per \cm}$, corresponding to the resonance condition for a detuning of $\Delta \omega_\text{exc}/2\pi \approx \qty{150}{MHz}$ from the zero-field Rydberg transition. This field magnitude is achieved at the cloud position ($z_0\sim25$µm) by superimposing an appropriate homogeneous bias field $E_\text{bias}$ onto the surface field. The two different bias field configurations, given by Eq.~\ref{eq:bias_fields}, that satisfy the resonant excitation condition are illustrated in Fig.~\ref{fig:TotalFieldBias}.
        
        \begin{figure}
            \centering
            \includegraphics[]{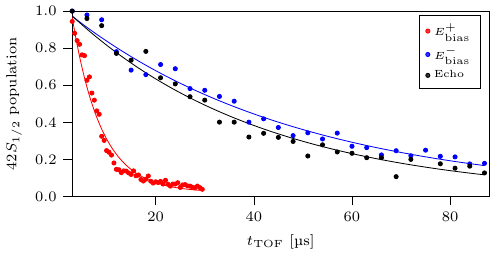}
            \caption{Rydberg state population after given $t_\text{TOF}$ as deduced from the total number of ion counts for $N_\text{rep}=1000$ excitation-detection cycles and normalized to the $t_\text{TOF}=0$ value. Atoms are excited at a total field magnitude of $|E_\text{res}| = \qty{4.45}{\V \per \cm}$. 
            Red: Field gradient accelerating atoms toward the chip, leading to rapid signal loss. Blue: Reversed field gradient accelerating atoms away from the chip and extending detection time. 
            Black: Application of the echo pulse sequence effectively cancels the Stark-induced position shift, preserving the Rydberg signal for durations approaching the natural lifetime. The solid lines are exponential decay fits yielding lifetimes of \qty{5.92}{\us}, \qty{47.12}{\us} and \qty{39.72}{\us} for red, blue and black respectively. The bias fields are $E_\text{bias} = \qty{-6.5}{\V\per\cm}$ and \qty{-12.8}{\V\per\cm} for red and blue, respectively, determined by the surface field with an additional lateral component of $E_{xy} \sim \qty{2.6}{\V\per\cm}$.}
            \label{fig:Lifetime}
        \end{figure}
        
        If the bias field is tuned such that the total field at the cloud position equals $+E_\text{res} = \qty{+4.45}{\V \per \cm}$ (corresponding to $E_\text{bias}^+$), but the overall field strength still decreases with distance from the chip, atoms experience a force towards the chip. In this configuration, atoms are rapidly lost from detection — either due to collisions with the surface or, more likely, because they move into regions where the surface electric field reduces the applied ionization field below the ionization threshold, rendering field ionization ineffective. This is visible in the red curve in Fig.~\ref{fig:Lifetime}.
        
        In contrast, when the bias field is increased such that the total field becomes negative ($-E_\text{res}$) at the cloud position, the magnitude of the total field increases with distance from the chip  (corresponding to $E_\text{bias}^-$). This reverses the field gradient at the cloud position and accelerates atoms away from the surface and towards the bias electrode. In this configuration, atoms remain in a low field region and are detected over significantly longer $t_{\text{TOF}}$ durations. The observed decay is shown as the blue curve in Fig.~\ref{fig:Lifetime}, leading to a lifetime of \qty{47.12}{\us}, slightly longer than the black-body-radiation limited lifetime at $T=\qty{300}{\K}$ of $\qty{40.9}{\us}$ \cite{Branden_Juhasz_Mahlokozera_Vesa_Wilson_Zheng_Kortyna_Tate_2009}, but lower than the natural Rydberg lifetime of \qty{\sim70}{\us} in the cryogenic environment, based on a quasiclassical calculation \cite{Beterov_Ryabtsev_Tretyakov_Entin_2009}.
        
        This difference in lifetime is consistent with the expectations, confirming that the observed signal decay is not caused by intrinsic state decay, but by motion through a spatially varying field. The key difference between these two cases is the sign of the field gradient, which determines the direction of the Stark force acting on the Rydberg atoms.

    \subsection{Stark echo}

        To experimentally demonstrate the effectiveness of the Stark echo scheme, we again perform time-of-flight measurements with de-excitation spectroscopy, this time combined with Stark echo pulses. The pulse sequence is illustrated in Fig.~\ref{fig:Scheme}(b) (bottom): atoms are first excited to the $42S_{1/2}$ state, evolve under an electric field with field gradient reversals, and are finally de-excited back to the ground state using a resonant laser pulse. Detection is performed via field ionization. Crucially, the atoms are always de-excited in the same electric field as they were initially excited, which ensures that any static Rydberg shift from the imperfect choice of bias fields is cancelled.
    
        Each data point in the Stark echo measurements corresponds to a full de-excitation spectrum, acquired in the same manner as shown in Fig.~\ref{fig:de-excitationSpectrum}. As before, the center of the spectrum measured by scanning the detuning of the de-excitation laser $\Delta \omega_\text{de-exc}$ provides a direct measure of the accumulated Stark shift experienced by the atoms during their time evolution. For the sake of simplicity, we used only a single echo pulse ($n=0$), but with an extra flip at $t_\text{TOF} = 0$ to ensure that the de-excitation is done in the same field as the excitation, such that $t_\text{TOF} = t_\text{echo}$.
    
        The flip of the electric field is implemented by switching between two bias voltages applied to the bias electrode below the atoms (see Fig.~\ref{fig:Scheme}(a)). These voltages are determined by performing Rydberg excitation spectroscopy as a function of applied bias field, as mentioned in Sec.~\ref{subsec:setup}, and then used as the two states for the square-wave voltage pattern applied after excitation. The switching signal is generated by our timing controller using an analog output channel, which is amplified and applied to the bias electrode. Current hardware limitations constrain the ramping time for the transition between the two voltages to $t_\text{ramp} \approx \qty{1}{\us}$. The performance of the echo scheme drastically drops if $t_\text{flip} < t_\text{ramp}$, which limits the minimal echo time to $t_\text{echo} > \qty{3.4}{\micro\second}$ for an effective echo sequence. 
        
        \begin{figure}
            \centering
            \includegraphics[]{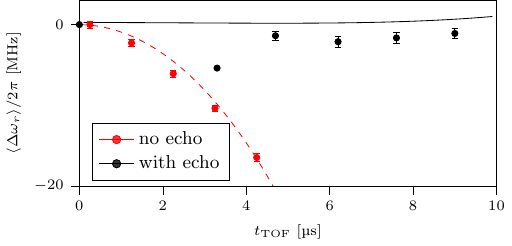}
            \caption{Average shift $\langle \Delta \omega_r \rangle / 2 \pi$ of the atomic resonance as a function of $t_{\text{TOF}}$ at $|E_\text{res}| = \qty{4.5}{\V \per \cm}$ ($\Delta \omega_\text{exc} / 2\pi \approx \qty{150}{\MHz}$). The red circles show measured data for $42S_{1/2}$ atoms in an inhomogeneous electric field without Stark echo sequence, exhibiting rapid frequency shift from resonance. The red dashed line is the corresponding simulation. The black circles show data for the case where a single Stark echo sequence is applied ($t_\text{TOF} = t_\text{echo}$), resulting in an almost complete suppression of the shift. The black solid line shows the simulated echo sequence, which reproduces the near-zero level shift. The error bars on the circles denote the 1$\sigma$ confidence interval of the fit.}
            \label{fig:MeasVsSim}
        \end{figure} 
    
        The results of the echo measurements are illustrated in Fig.~\ref{fig:MeasVsSim}, showing the measured average level shift $\langle \Delta \omega_r \rangle / 2 \pi$ of the de-excitation resonance as a function of $t_{\text{TOF}}$. The red circles correspond to experiments without the echo sequence, resulting in a rapid growth of redshift, with level shifts exceeding $\qty{15}{\MHz}$ after just a few microseconds, in good agreement with our model (red dashed line).
    
        With the Stark echo applied (black circles), even a single echo cycle with $t_\text{TOF} = t_\text{echo}$ almost entirely suppresses the level shift. The measured shift $\langle \Delta \omega_r(t_\text{TOF}) \rangle / 2 \pi$ for $t_\text{echo}>\qty{3.4}{\micro\second}$ remains within a range of \qty{1}{\MHz} over the full measurement with only a small offset of $\approx \qty{-1.5}{\MHz}$. The corresponding simulation (assuming instantaneous switching) reproduces this behavior and yields only minor residual shifts due to the inhomogeneity of the electric field (solid black line). The near-complete cancellation of the level shift confirms the effectiveness of the echo sequence in suppressing motional dephasing. The offset in the measurement data is most likely caused by an imperfect setting of $E_\text{bias}$ and timing imperfections on the \qty{100}{\ns} scale. The data point at $t_\text{TOF} = \qty{3.3}{\us}$ corresponds to $t_\text{flip} \approx t_\text{ramp}$ above which the echo sequence starts becoming effective. An echo sequence with four field switches ($n=2$) was also measured for $t_\text{TOF} \approx \qty{11}{\us}$, showing comparable results to the single echo sequence of the same length.
    
        The effectiveness of the echo sequence is further confirmed by lifetime measurements, shown in Fig.~\ref{fig:Lifetime}. When the echo sequence is applied (black points), the measured Rydberg lifetime remains consistent with the natural radiative lifetime. Together, these measurements demonstrate that the Stark echo technique effectively suppresses both level shift and signal loss due to field-induced atomic motion.

\section{Conclusion and Outlook}\label{sec:conclusion}
    We have investigated the motion of Rydberg atoms caused by inhomogeneous electric fields near a superconducting atom chip surface. The spatially varying stray fields induce position-dependent Stark shifts and forces that accelerate the atoms, leading to time-dependent Rydberg level shifts. Through a combination of de-excitation spectroscopy, lifetime measurements, and numerical simulations, we characterized this effect and demonstrated that even modest polarizabilities lead to significant level shifts on microsecond timescales.

    To mitigate this effect, we introduced a Stark echo scheme based on dynamically reversing the electric field. By alternating the direction of the Stark force, atomic motion and accumulated level shifts are effectively cancelled. We implemented the echo sequence in a chip-based Rydberg platform and observed a clear recovery of the resonance, consistent with suppression of the motion. This technique requires only global field control and is compatible with a wide range of trapping geometries and experimental protocols.
    
    The method was shown to significantly reduce the Rydberg level shift in our experimental platform where our broader motivation is to couple Rydberg atoms to an on-chip CPW resonator, where strong atom–resonator coupling necessitates trapping the atoms within tens of microns from the chip surface. In this regime, stray electric fields arising from adsorbates, patch potentials, or surface charges are difficult to suppress entirely and often lead to uncontrolled gradients. The Stark echo sequence provides a robust and flexible approach for suppressing the resulting motion and decoherence without interfering with the microwave coupling, as only the direction — not the magnitude — of the electric field is reversed during the sequence.
    
    Beyond suppressing motion in the longitudinal ($z$) direction, the method can also be extended in principle to address residual electric field components in the transverse ($x$/$y$) directions. Above local adsorbate patches, the lateral field components are typically much weaker than the longitudinal component, with correspondingly smaller gradients. While direct compensation of these transverse fields over spatially extended regions may not be feasible with a global bias, further mitigation may be achieved via filtering techniques by exploiting the narrow linewidth of the excitation and de-excitation lasers. By aligning them along transverse directions, only atoms within a narrow velocity class can be coherently driven, suppressing the influence of fast-moving atoms.

    In addition, the Stark echo scheme inherently acts as a velocity filter. Atoms that acquire significant velocity due to strong local gradients no longer return to their initial state and can be preferentially removed from the cloud. This opens the possibility of implementing a selective cooling mechanism: Rydberg atoms that cannot be recaptured by the echo sequence are effectively expelled, gradually narrowing the motional distribution of the ensemble. Such echo-based cooling schemes may prove useful for preparing low-entropy samples in surface-based quantum devices.
\section*{Acknowledgements}
This work was supported by the Deutsche Forschungsgemeinschaft (DFG) through the Research Unit FOR 5413 (Grant No. 465199066) and the QuantERA Programme (MOCA-Project No. 491986552).

\appendix
\end{document}